\documentclass[10pt, conference, compsocconf]{IEEEtran}
\def\MeV {\,{\tt MeV}}

\def\GBs {\,{\tt GB/s}}
\def\GB {\,{\tt GB}}
\def\GHz {\,{\tt GHz}}
\def\GFlops {\,{\tt GFlops/s}}

\def\sign {\mathop{\hbox{sign}}}

\usepackage{dsfont}  
\usepackage{bm}      
\def\iden{\mathds{1}}

\def\beq{\begin{equation}}
\def\eeq{\end{equation}}

\def\comment#1{}

\usepackage{slashed}
\newcommand{\dslash}{\slashed{D} }

\ifCLASSINFOpdf
   \usepackage[pdftex]{graphicx}
\else
\fi
\usepackage[cmex10]{amsmath}

\usepackage[colorlinks=true,backref=false,linktocpage=true,
citecolor=blue,urlcolor=blue,linkcolor=blue,pdfpagemode=UseOutlines]{hyperref}

\begin{document}
%
\title{Efficient implementation of the overlap operator on multi-GPUs}


\author{\IEEEauthorblockN{Andrei Alexandru, Michael Lujan, Craig Pelissier, 
Ben Gamari, Frank Lee}
\IEEEauthorblockA{Department of Physics,
The George Washington University\\
725 21$^\text{st}$ St. NW, 
Washington, DC 20052}
}


%


\maketitle

\begin{abstract}
Lattice QCD calculations were one of the first applications to show
the potential of GPUs in the area of high performance computing. 
Our interest is to find ways to effectively use GPUs for lattice
calculations using the overlap operator. The large memory footprint of these
codes requires the use of multiple GPUs in parallel. In this paper we show
the methods we used to implement this operator efficiently. We run our codes
both on a GPU cluster and a CPU cluster with similar interconnects. We find
that to match performance the CPU cluster requires 20-30 times
more CPU cores than GPUs.
\end{abstract}


%
\IEEEpeerreviewmaketitle

\section{Introduction}

The structure of subnuclear particles like the proton and neutron is dictated by the 
dynamics of quarks and gluons. The {\em strong force} that binds them is described 
using quantum chromodynamics (QCD). Lattice QCD is a discretized version of this theory
that makes it amenable to numerical simulations. The calculations involved are
very demanding but they can be parallelized efficiently. Consequently, most lattice
QCD simulations are run on traditional CPU clusters using fast interconnects.
A recent alternative is to use graphics processing units (GPUs) for lattice QCD
simulations~\cite{Egri:2006zm,Clark:2009wm,Babich:2010mu,Alexandru:2011ee}.
While difficult to program, these devices have very good floating point 
performance and very good memory bandwidth. For lattice QCD simulations,
GPUs can outperform CPUs by large factors which allow us to build
more compact and cost effective clusters. Currently most lattice QCD
simulations run either in single GPU mode or on {\em fat nodes} with a few GPUs
communicating over the PCI bus. This is the most efficient configuration
if the memory requirements are relatively modest.
However, this is not always feasible.

Lattice QCD simulations can be performed using different discretizations of the QCD action
depending on the problem studied. Traditional discretizations for the quark field
introduce artifacts that are removed only in the {\em continuum limit}, i.e. when lattice
spacing goes to zero. In particular, they break {\em chiral symmetry} which plays
an important role for simulations close to the physical limit. {\em Overlap} 
discretization~\cite{Neuberger:1997fp} of the quark field preserves this 
symmetry which allows us to capture the effects of chiral dynamics even at 
finite lattice spacing. 

Overlap formulation is numerically demanding and an efficient implementation
requires significantly more memory than traditional discretizations. To implement
it on GPUs we need to be able to break the problem on multiple GPUs in order
to satisfy the memory requirements. In this paper we present our implementation
of the overlap operator in multi-GPU context. The outline of the paper is
the following. In Section~\ref{sec:overlap} we review the numerical properties 
of the overlap operator. In Section~\ref{sec:parallel} we discuss our parallelization 
strategy and the GPU-GPU communication structure. In Section~\ref{sec:dslash} we 
discuss the implementation of the {\tt dslash} routine, which is the building block 
for the overlap operator. In Section~\ref{sec:implementation} we discuss 
the implementation 
of the overlap operator, the required eigensolvers and conjugate gradient (CG) inverter 
used to compute the quark propagators.

\section{Overlap operator}\label{sec:overlap}

\begin{figure}[b] 
\centering 
\includegraphics[width=3.2in]{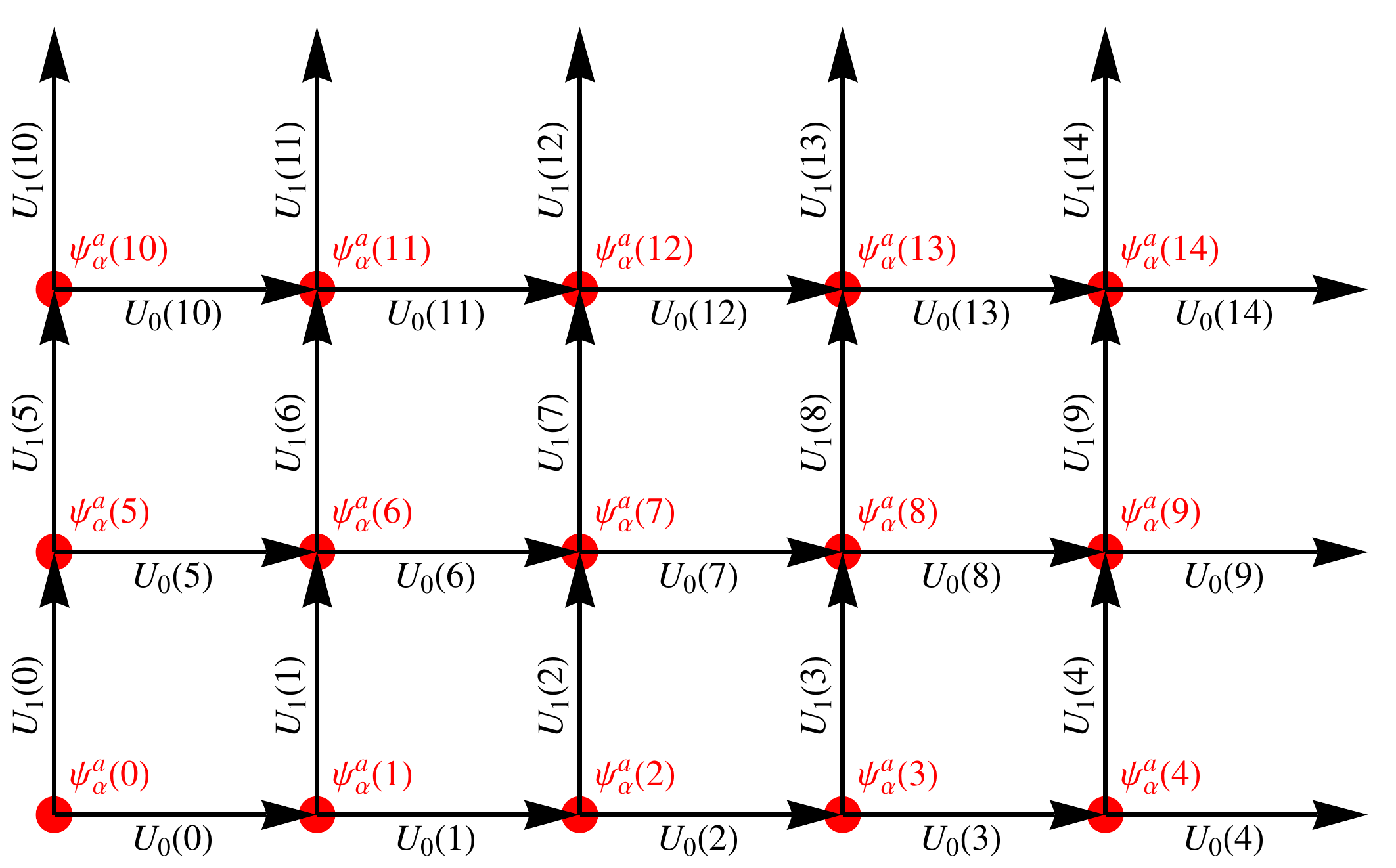} 
\caption{Schematic representation
of the lattice discretization: the quark fields $\psi^a_\alpha$ are
associated with the sites of the lattice and the gauge variables
$U_\mu$ are defined on the links connecting the sites.} 
\label{fig:lattice}
\end{figure}

In lattice QCD the space-time is approximated
by a four dimensional grid, the quarks are viewed as particles
hopping between the grid sites and gluons are represented
by parallel transporters that change the internal state of the
quarks as they hop along the given link (see Fig. \ref{fig:lattice}
for a schematic representation). The quark operator represents
a discretization of the covariant derivative $D_\mu = \partial_\mu + i g A_\mu$,
where $A_\mu$ is the color (gluon) field. The quark fields, $\psi_n$,
are represented by $4\times 3$ matrices at each lattice site and the gluon field $U_\mu(n)=e^{iag A_\mu(n)}$ by $SU(3)$ matrices. 
Wilson discretization of $m+\dslash = m+\gamma_\mu D_\mu$ is given by the following matrix
\beq
\label{eq:1}
D_w(m; U)=(m a+4)\iden-\frac{1}{2} \sum_{\mu=\pm1}^{\pm4} T_{\mu}(U) \,,
\eeq
where $T_\mu$ are the {\em parallel transporters} for all 8 directions
\beq
\label{eq:2}
\begin{split}
\mu>0:\quad(T_\mu \psi)_{n} &= U_\mu(n) \psi_{n+\hat\mu} (1-\gamma_\mu) \,,\\ 
\mu<0:\quad
(T_\mu \psi)_{n} &= U_\mu(n-\hat\mu)^\dagger \psi_{n-\hat\mu} (1+\gamma_\mu) \,.
\end{split}
\eeq
$D_w$ is a complex $12 N \times 12 N$ matrix, where $N$ is the number of sites 
on the grid. The matrix is very sparse since the parallel transporters only connect
to nearest-neighbor sites. For numerical simulations we store only the non-zero elements
of this matrix and we implement a {\tt dslash} routine to compute $D_w\psi$ on 
any given quark field $\psi$. The storage requirements and the numerical cost
for the {\tt dslash} routine are proportional to $N$. 
$D_w$ is $\gamma_5$-symmetric, i.e. 
$D_w^\dagger = \gamma_5 D_w \gamma_5$. Thus, $H_w\equiv\gamma_5 D_w$ is
hermitian.

The massless overlap operator is defined in terms of the Wilson operator
\beq
D = \iden + \gamma_5 \sign(H_w) \,.
\eeq
As in the case of Wilson fermions, for numerical studies we need to implement
a routine that computes $D\psi$ for any quark field $\psi$. This calls for a 
practical algorithm to compute the matrix sign function, i.e. $\sign(H_w)\psi$. 
This can be done using either a polynomial approximation for the sign function
\cite{Giusti:2002sm} or a rational approximation \cite{Chiu:2002eh}. In both 
cases an optimal approximation, $P(x)$, for $x^{-1/2}$ is determined such that
\beq
\delta=\max_{x\in[\epsilon,1]} \left| 1-\sqrt x P(x) \right|\,,
\eeq
is minimized over the set of functions used. $P(x)$ is either a polynomial 
$p_0+p_1 x + \dots + p_n x^n$ of order $n$ or a rational function 
$(p_0 + p_1 x + \dots + p_{n-1} x^{n-1})/(q_0 + q_1 x + \dots + q_n x^n)$.
The coefficients of the polynomial approximation can be determined using
a robust numerical method~\cite{Giusti:2002sm} and the coefficients of the
optimal rational approximation have been determined analytically~\cite{Chiu:2002eh}.

\begin{figure}[t]
\centering
\includegraphics[width=2.8in]{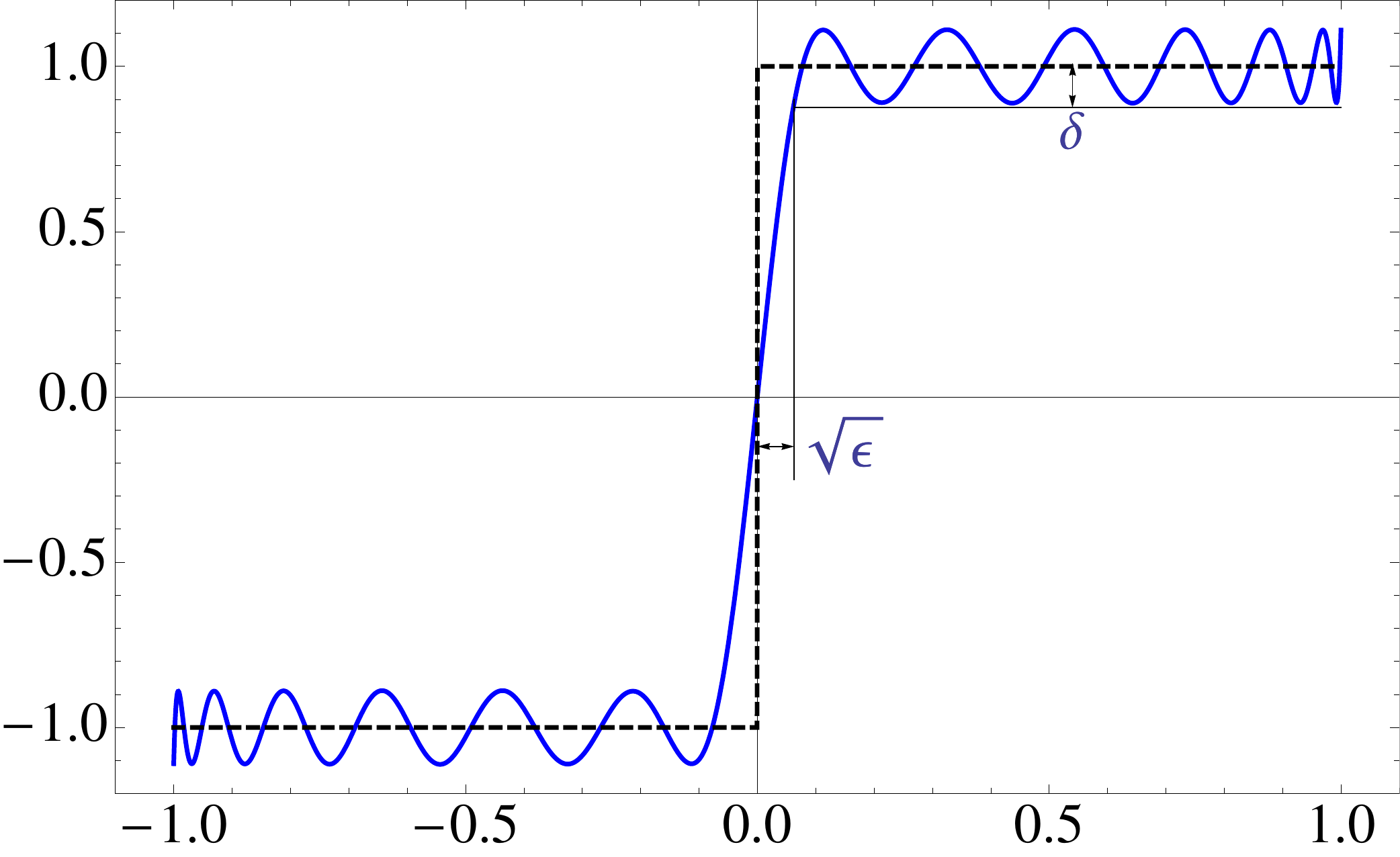} 
\caption{Polynomial approximation $xP(x^2)$ for the sign function with 
$\delta=0.1$, $n=13$ and $\sqrt\epsilon\approx0.06$.} 
\label{fig:sign-approx}
\end{figure}

Using this result, we can approximate the sign function using
\beq
\sign(H_w) \approx Q P(Q^2) \quad\text{with}\quad 
Q=\frac{H_w}{ \left\Vert H_w \right\Vert}\,.
\eeq
A small order approximation is presented in Fig.~\ref{fig:sign-approx}.
Note that the approximation is quite poor for the interval 
$x\in[-\sqrt\epsilon,\sqrt\epsilon]$. This is a generic feature: a
continuous odd function will go through $0$ at $x=0$. Thus, there is
always a neighborhood around $x=0$ where the approximation is poor.
To shrink this region we have to increase the order of the approximation.
Since the approximation needs to be good over the whole spectrum of $Q$, the size
of this region has to be adjusted such that $\sqrt\epsilon < |\lambda_\text{min}|$,
where $\lambda_\text{min}$ is the eigenvalue of $Q$ closest to zero.
For typical lattices we often have $|\lambda_\text{min}|\sim 10^{-4}$ and
to get an approximation of the order $\delta=10^{-10}$ we would need 
polynomials of the order $n\sim 10^5$ -- this is impractical. 

The standard
solution is to determine the spectrum of $Q$ in the neighborhood of zero and
use it to exactly calculate the sign function in this subspace. The approximation
is then needed only on the orthogonal subspace, and on this subspace 
small order polynomials lead to very good approximations. A similar argument
leads to the same conclusion for the rational approximation.

To be more specific,
assume we have determined the $\ell$ eigenmodes of the $Q$ matrix closest 
to zero, $Q \eta_i = \lambda_i \eta_i$ ordered such that 
$|\lambda_1|<|\lambda_2|<\dots<|\lambda_\ell|$. We can then compute
$\sign(H_w)\psi$ by separating $\psi = \psi_l + \psi_h$ with the low 
frequency component defined by 
\beq
\psi_l \equiv P_l \psi = \sum_{i=1}^\ell \eta_i^\dagger \psi \:\eta_i \, .
\eeq
We have then
\beq
\begin{split}
\sign(H_w)\psi &= \sign(H_w)\psi_l + \sign(H_w)\psi_h\\
&= \sum_{i=1}^\ell \sign(\lambda_i) \eta_i^\dagger\psi \: \eta_i + \sign(H_w)\psi_h\\
&\approx \sum_{i=1}^\ell \sign(\lambda_i) \eta_i^\dagger\psi \: \eta_i + Q P(Q^2)\psi_h \,.
\end{split}
\eeq
The advantage is that the approximation $Q P(Q^2)\psi_h$ is good if 
$\sqrt\epsilon = |\lambda_\ell|$. For typical lattices using $\ell\sim 100$
produces $|\lambda_\ell|\sim 10^{-1}$ which can be approximated using
polynomials with $n\sim 100$.

The overlap operator for quarks of mass $m$
\beq
D(m) = \rho D+ma(1-\frac12 D),
\eeq
is used to compute the quark propagators $(1-\tfrac12 D)D(m)^{-1}\psi$. 
We use a multi-shifted version of CG~\cite{Jegerlehner:1996pm}
to compute the propagators for multiple masses at once.
The conditioning number for this matrix is very large for quark masses
close to physical values and we need to use {\em deflation} to accelerate
convergence~\cite{Li:2010pw}. Take the $\ell'$ eigenmodes 
of the massless overlap operator closest to zero, i.e. $D\xi_i=d_i\xi_i$,
with $|d_1|<|d_2|<\dots<|d_{\ell'}|$, we have
\beq
D^{-1}(m)\psi = \sum_{i=1}^{\ell'} \frac{ \xi_i^\dagger\psi}
{\rho d_i+ma(1-\frac12 d_i)}\xi_i + D(m)^{-1}\psi_h,
\eeq
where the high-frequency part of the propagator $D(m)^{-1}\psi_h$ converges at 
a rate controlled by the effective conditioning number 
$\kappa'=\Vert D \Vert / |d_{\ell'}|$ rather than $\kappa = \Vert D \Vert / ma$.

In order to compute quark propagators for the overlap operator efficiently we 
need to store the $\ell$ eigenvectors 
of the $H_w$ operator and the $\ell'$ eigenvectors of the $D$ operator in memory. 
Since $\ell,\ell'\sim 100$, the memory requirements for these codes are ten to 
a hundred times greater than the ones required for traditional discretizations.
This is the reason to implement these routines to run in parallel on multiple GPUs.
To carry out lattice QCD simulations using overlap discretization
in an efficient manner, we need to implement the following routines:
\begin{itemize}
\item[-] {\tt dslash} routine to compute $D_w\psi$,
\item[-] eigensolver to compute the lowest eigenmodes of $H_w$,
\item[-] massless overlap multiplication routine to compute $D\psi$,
\item[-] eigensolver to compute the lowest eigenmodes of $D$,
\item[-] multi-shifted CG to compute $D(m)^{-1}\psi$.
\end{itemize}
All these routines can be parallelized efficiently. Our strategy is described
in the next section.

\section{Parallelization strategy}\label{sec:parallel}

\begin{figure}[b]
\centering
\includegraphics[width=3.0in]{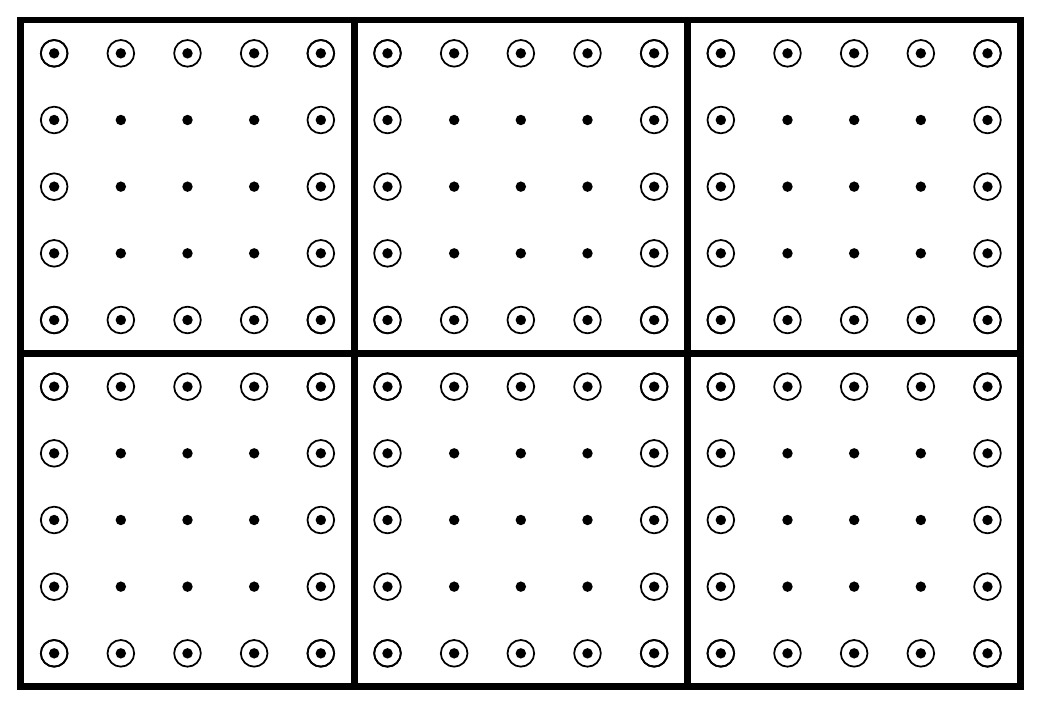}
\caption{Two dimensional $15\times 10$ lattice divided in 6 equal pieces. The
circled points require off-node data for the {\tt dslash} routine.}
\label{fig:div-lattice}
\end{figure}

The linear algebra algorithms used to compute the eigenvectors (implicitly
restarted Arnoldi~\cite{Sorensen:1992fk,Lehoucq:1996}) and the inverse of 
the lattice operators (CG) are not easy to parallelize. 
In fact in our codes these algorithms are executed in lockstep 
on all nodes. However, it turns out 
that most of the computational time is spent computing the matrix-vector 
multiplication $\phi \gets D_w\psi$ and the vector operations, e.g. 
$\phi \gets \alpha\psi_1 + \beta\psi_2$. The complexity of these routines
scales with the lattice volume but they can be efficiently parallelized.
The key feature is that these routines are {\em local}, i.e. the calculations
required to derive the value of $\phi$ at a given lattice site need only
the values of the source fields $\psi$'s at neighboring sites.

This suggests the optimal strategy for parallelization: we divide
the lattice in equal regions. A schematic representation of the
procedure is presented in Fig.~\ref{fig:div-lattice}. Each parallel 
{\em process} is responsible for one of the regions: all data
that are associated with the sites in this region resides in the
memory managed by this process and all calculations related to
these sites are carried out by this process. The lattice is 
divided in regions that have the same shape so that the parallel
processes run through the same steps, effectively using
a single-instruction, multiple-data (SIMD) paradigm.

\begin{figure}[t]
\centering
\includegraphics[width=3.2in]{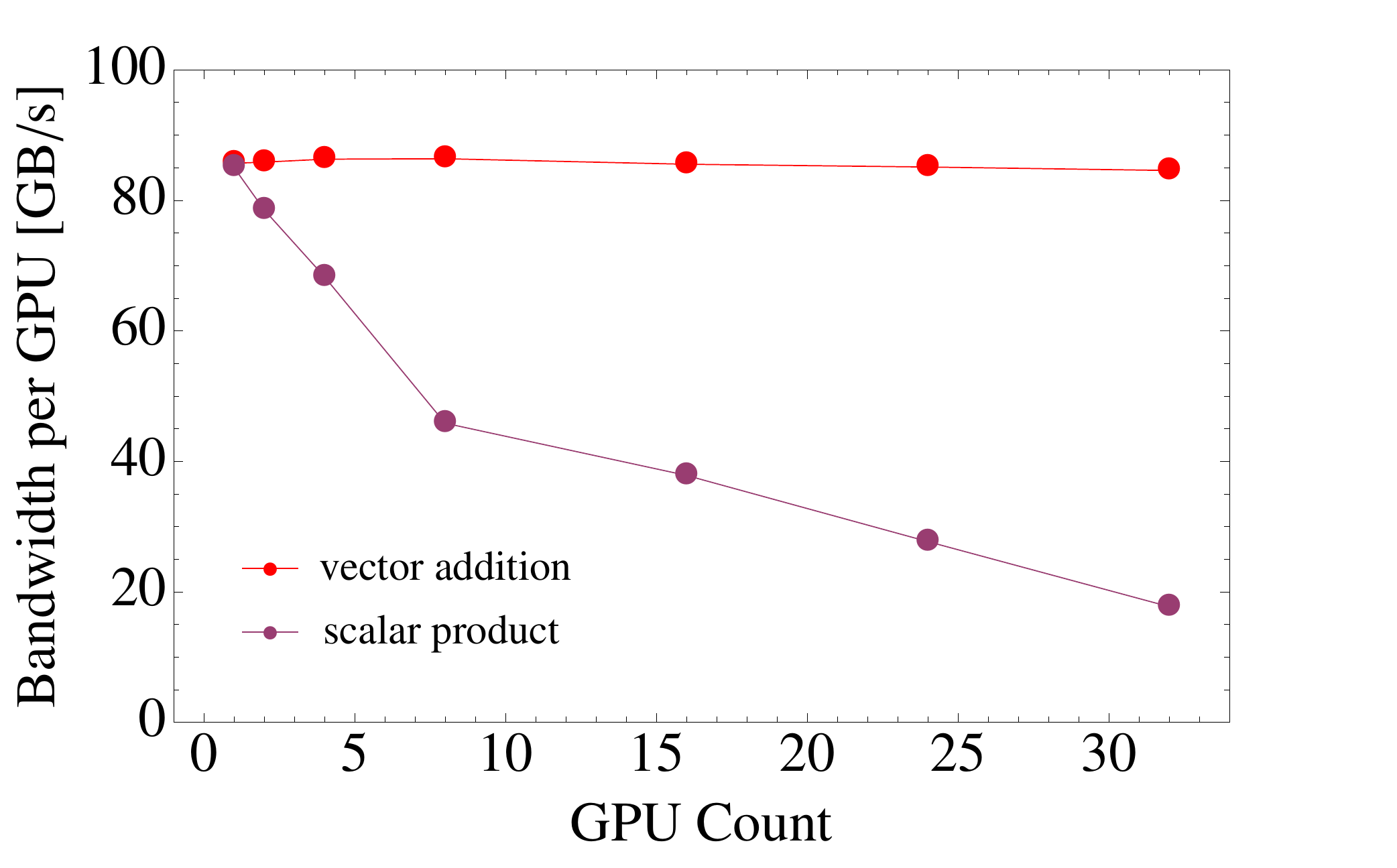}
\caption{Strong scaling on a $24^3\times 64$ lattice for vector operations; horizontal
lines in this plot indicate perfect scaling.}
\label{fig:vecop-scaling}
\end{figure}

For routines that are completely
local, i.e. routines where the result $\phi(s)$ at a site $s$ involves
only data associated with the same site, each process can 
proceed independently of the other processes. These routines
should scale perfectly when we divide the workload over multiple
processes. In Fig.~\ref{fig:vecop-scaling} we show the scaling
plot for vector operations as we increase the number of parallel
tasks. The relevant measure for vector operations is the effective 
memory bandwidth since this is the bottleneck. All purely local vector
operations use the same bandwidth as vector addition and show perfect scaling.

Vector operations that involve {\em reductions}, the scalar product
and vector norm, scale rather poorly. However, the poor scaling is
not a result of inter-process communication. Rather, as we divide the
lattice into smaller and smaller pieces, each GPU operates on smaller vectors
and the scalar product routine runs less efficiently~\cite{Alexandru:2011ee}.
When running a $24^3\times64$ lattice on 32 GPUs, each process is responsible
for a region of size $12^3\times16$. If we run a single-GPU scalar product on
a $12^3\times16$ lattice the bandwidth is $26.5\GBs$, compared to $18\GBs$
as measured when running $24^3\times64$ lattice on 32 GPUs. Fortunately, the
scalar products are responsible only for a small fraction of the computational
cost and their poor scaling has little impact overall.

For routines like
{\tt dslash} where the result $\phi(s)$ depends on the values
of neighboring site, inter-process communication is required.
This is only needed for sites that are adjacent to the borders,
the sites represented by circled points in Fig.~\ref{fig:div-lattice}.
When the communication cost is small compared with the computation
cost, the calculation can be efficiently run in parallel.

The division strategy we used is designed to minimize the number of 
sites on the boundary. This is based on the assumption that the
communication time will be proportional to the boundary area.
This assumption is valid when the communication speed between 
any two processes is the same.
For heterogenous systems, where the network ``distance" between 
processes varies, this division strategy might not be optimal. 
The results presented in this paper were produced on homogenous systems.

Each process is represented by a CPU thread attached to a single GPU.
In the implementation discussed here the most compute intensive routines
are executed on the GPU. For maximum performance we place all
frequently accessed data in GPU memory. The GPU {\em kernels} were
coded using CUDA~\cite{CUDA}.
We use MPI~\cite{MPI-2.2} to communicate between processes.
MPI calls send/receive buffers stored in CPU memory. To send
data stored in GPU memory the sender process first copies the buffer to 
its CPU memory, and then uses regular MPI calls to send it to the 
appropriate process. After receiving the data, the target process copies
it to its GPU memory. Thus, sending data between GPUs involves two additional
transfers over the PCI bus. This adds a significant overhead since the PCI bus
is not much faster than the interconnects used in our test systems. 
To alleviate this problem we use {\em pinned} memory for the CPU
communication buffers so that GPU to CPU data transfers run at full PCI bus
speed. Close attention needs to be paid in handling these buffers since
they are a scarce resource. Another issue we encountered is that OpenMPI~\cite{OpenMPI} 
Infiniband library uses pinned memory in an incompatible manner --  
this conflict is resolved by turning off the pinned memory optimization for 
this MPI library.

\section{Multi-GPU {\tt dslash} implementation}\label{sec:dslash}

The most time consuming step in our codes is the {\tt dslash} routine that
computes $\phi\gets D_w\psi$. Consequently, our optimization efforts
focused primarily on implementing it efficiently. We are primarily interested
in the double precision implementation since the eigensolvers employed
are very sensitive to roundoff errors. For the most part, we used
the optimizations, data layout and codes developed for our single-GPU 
implementation~\cite{Alexandru:2011ee}. The calculation of $\phi$ for the
{\em bulk} sites, the sites that do not require off-process data, is carried
out using exactly the same steps. To deal with the boundary sites, we need
to implement:
\begin{itemize}
\item[-] a {\tt gather} routine that performs the required calculations
and collects the data to be sent off-process in communication buffers, 
\item[-] a {\tt scatter} routine that moves the data from the communication
buffers to the appropriate sites and performs the required color multiplications,
\item[-] communication routines that copy the GPU buffers from sender to receiver.
\end{itemize}
The {\tt scatter}/{\tt gather} routines add very little overhead. Moving the
data from one GPU to another is the most time consuming step. Fortunately, this
step can be carried out in parallel with the bulk {\tt dslash} calculation.
As discussed in the previous section, the GPU to GPU communication occurs in 
three steps: GPU to CPU, CPU to CPU and CPU to GPU. The CPU to CPU transfer can
be executed using a non-blocking MPI call. The CPU to GPU data transfer can
be performed in parallel with the {\tt dslash} kernel only if we use 
asynchronous CUDA copy instructions~\cite{CUDA}. This only works if the CPU
buffer involved uses pinned memory. 

\begin{figure}[t]
\centering
\includegraphics[width=2.8in]{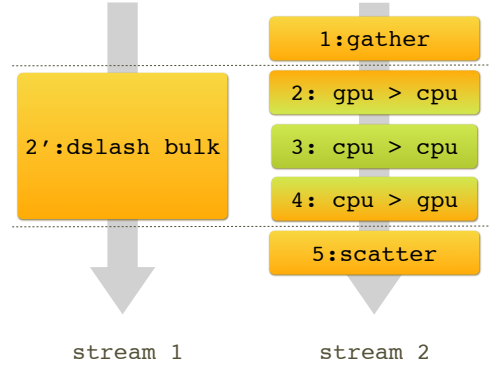}
\caption{Schematic diagram of the scheduling order for the multi-GPU
{\tt dslash} routine. The dashed lines represent CUDA synchronization points.}
\label{fig:dslash-order}
\end{figure}

Since the GPU kernels are issued
asynchronously, we need to arrange carefully the execution order to 
ensure logical consistency. To achieve this we had to use CUDA {\em streams}:
kernels attached to a particular stream are guaranteed to be executed in the
order issued. Kernels, or asynchronous CUDA copy instructions, can be
executed in parallel if they belong to different streams. The logical
structure of the multi-GPU {\tt dslash} routine is represented schematically
in Fig.~\ref{fig:dslash-order}. Since {\tt gather}, bulk {\tt dslash}, and 
{\tt scatter} kernels have to be executed sequentially, although they are
attached to different streams, CUDA synchronization calls are used to enforce
this constraint.

A parallel code is efficient when the aggregate performance is proportional
to the number of processes. We present here the results for {\em strong} scaling,
i.e. performance of our codes for a lattice of fixed size that gets divided
into smaller and smaller pieces as the number of processes is increased. Since
the {\tt gather}/{\tt scatter} kernels take very little time, as long as the
bulk {\tt dslash} kernel takes more time than communication, the scaling
will be almost perfect. However, as we increase the number of GPUs, the time
for the bulk {\tt dslash} kernel will decrease as $N_\text{GPU}^{-1}$, whereas
the communication time will only decrease as $N_\text{GPU}^{-3/4}$.
Eventually the communication time will dominate and scaling will suffer.

\begin{figure}[t]
\centering
\includegraphics[width=3.2in]{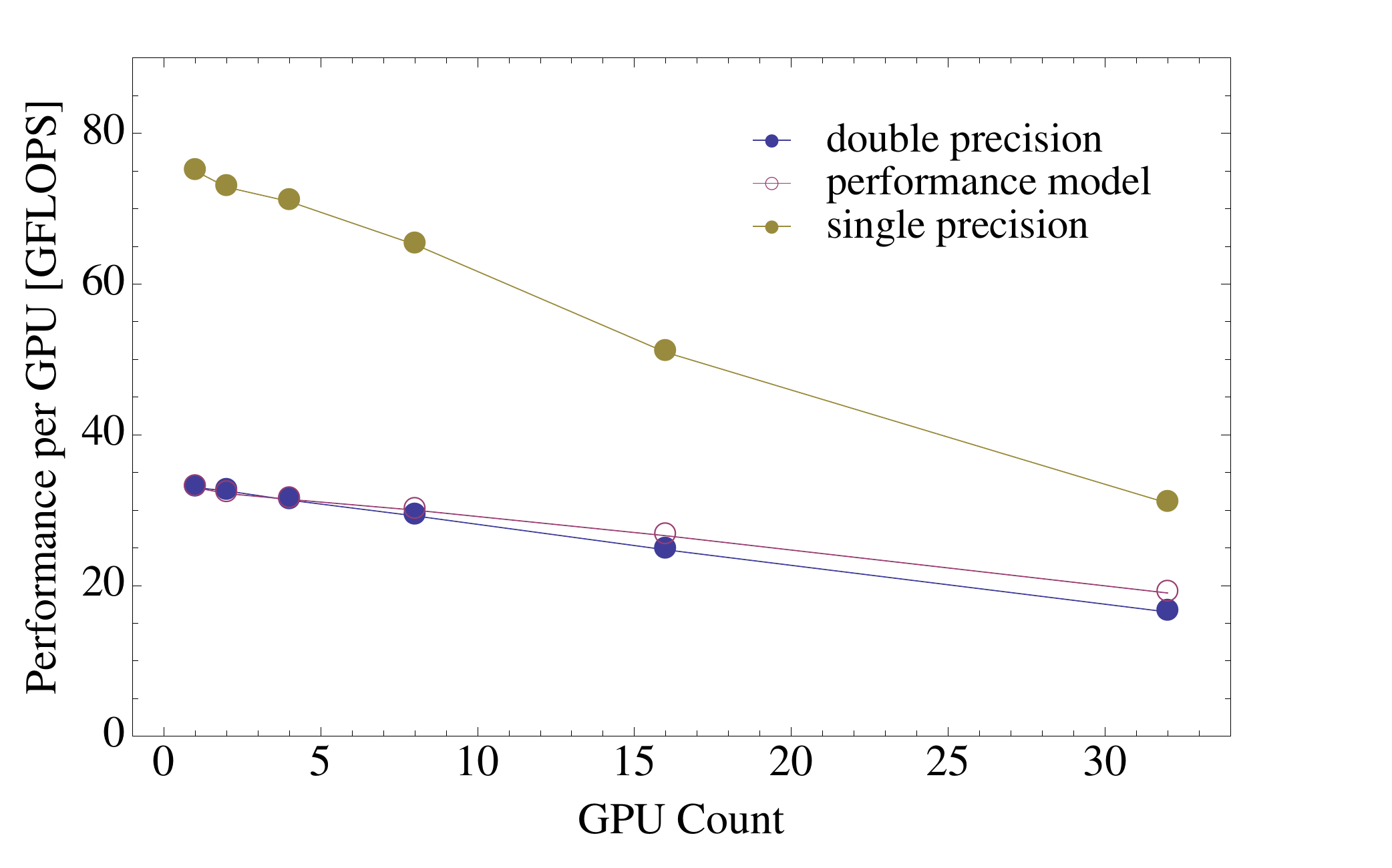}
\caption{Strong scaling for the multi-GPU {\tt dslash} on a $24^3\times 64$ lattice.
The empty symbols represent the results of a performance model for the double
precision routine and the solid points are measured on the GPU cluster described
in the text.}
\label{fig:dslash-performance}
\end{figure}

The results presented here are measured on a GPU cluster with a single Tesla M2070 per
node using QDR Infiniband network. On GPUs the error correction mode 
is turned on; this ensures the accuracy of the result at the expense of reducing 
the memory bandwidth. Since our kernels are memory bound their performance is
also affected. The bandwidth measured in micro-benchmarks are $3.2\GBs$ for
the PCI bus and $4.2\GBs$ for Infiniband. The {\tt gather}/{\tt scatter} kernels
move 45/69 numbers per boundary site and their bandwidth performance is $55\GBs$.
The double precision bulk {\tt dslash} kernel performance is about $33\GFlops$.
Using these numbers we can construct a simple performance model for the multi-GPU
routine. In Fig.~\ref{fig:dslash-performance} we present the performance of our
codes and compare it with the predictions from the performance model. We see that
for 32 GPUs our scaling efficiency is about $50$\% for a $24^3\times 64$ lattice.
For larger lattices the scaling is even better, for example for a $32^3\times 64$
lattice the 32 GPUs scaling efficiency is about $60$\%.
Our simple model follows closely the measured results for the double precision routine,
but it overestimates slightly its performance. This is most likely due to the fact that,
as in the case of the scalar product discussed in the previous section, the efficiency
of the GPU kernels decreases as the size of the vectors gets smaller.

\begin{figure}[t]
\centering
\includegraphics[width=3.2in]{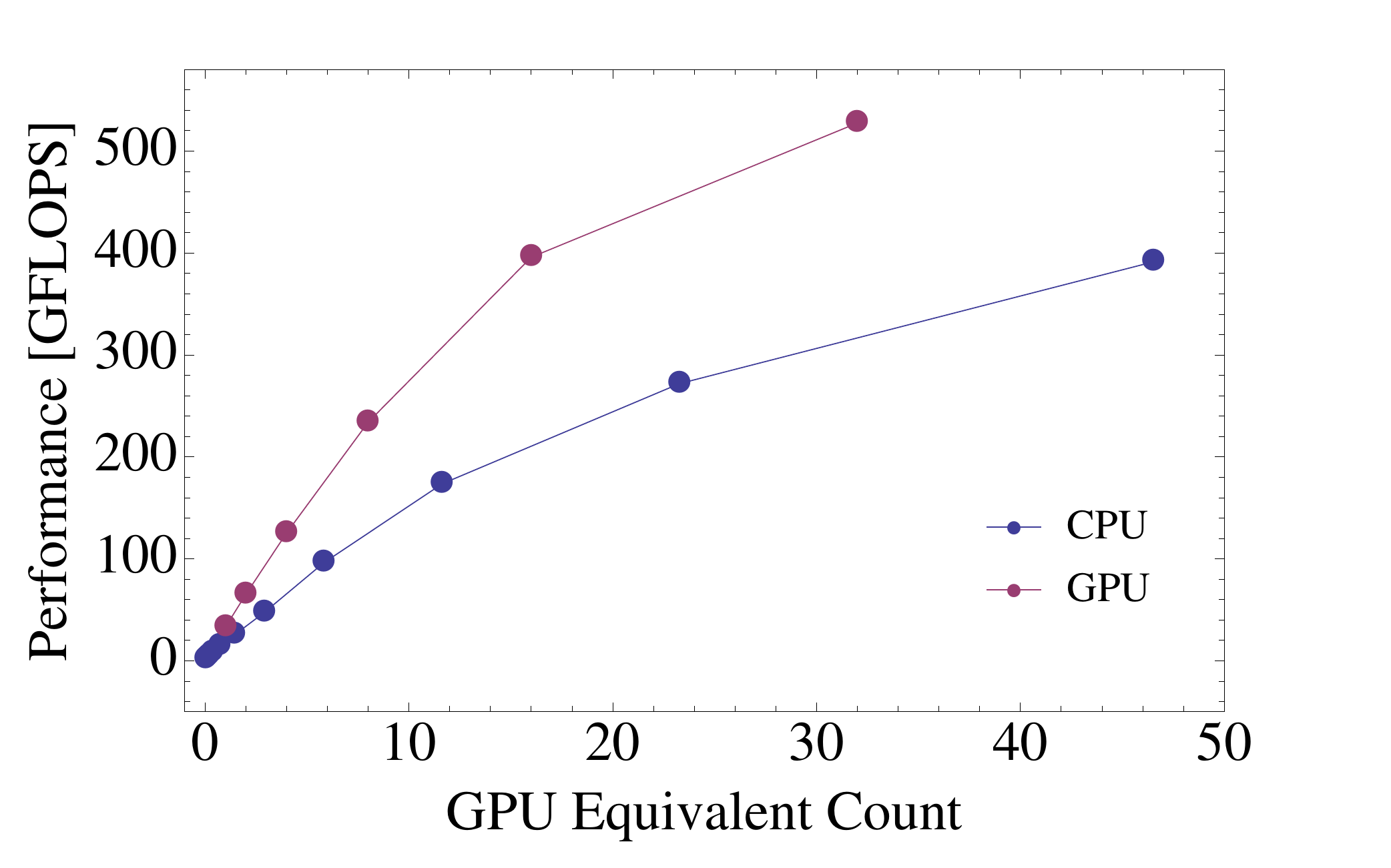}
\caption{Strong scaling for the multi-GPU and multi-CPU {\tt dslash} on a $24^3\times 64$ lattice. The CPU-core count is rescaled by a factor of 22.}
\label{fig:dslash-comp}
\end{figure}

To get a better picture, it is instructive to compare the performance of the GPU
code with an equivalent code running purely on CPUs. The typical CPU {\tt dslash}
performance for double precision implementations is 
$1\text{--}2\GFlops$~\cite{Wettig:2005zc}. Our own CPU implementation runs at
$1.5\GFlops$ per core. This number was measured on a Cray XT-5 machine that uses
very fast interconnects and dual hex-core AMD CPU per node. The CPUs run at 
$2.6\GHz$. When comparing single-GPU performance it is then easy to see that
the performance of one GPU is equivalent to 22 CPU cores. In the multi-GPU
context it is less straightforward to define a measure, since scaling also 
plays an important role. To aid this comparison, we carried out a strong
scaling study using our CPU {\tt dslash} implementation running on the Cray
machine. To compare the GPU and CPU performances we plot the aggregate performance
of both CPU and GPU codes in Fig.~\ref{fig:dslash-comp}: the CPU core count
is translated into its GPU equivalent by dividing the total number of CPU cores by 22.
This insures that the leftmost points in the graph overlap. If the CPU code
scales similarly to the GPU code, the two curves should overlap. It is clear that
the GPU codes scale better and that the equivalent CPU core count increases
as we increase the number of GPUs. For example, the aggregate performance
for 32 GPUs is $527\GFlops$ whereas the performance of $32\times 22$ CPU-cores
is only about $300\GFlops$.

\section{Overlap operator implementation}\label{sec:implementation}

Using the {\tt dslash} and vector routines discussed in the 
previous sections we build now the overlap operator. 
To present the performance of our codes, we employ $24^3\times 64$ lattices
from one of our lattice QCD projects. The GPU cluster used
for our testing has 32 GPUs with $3\GB$ of memory per GPU. The total GPU memory is 
then only sufficient to accommodate about $500$ vectors. To compare the GPU
performance with CPU codes, we run similar calculations on the Cray XT-5
machine described in the previous section. Since the scaling of the CPU
codes is poorer than our GPU codes, we run our codes on 256 cores which
is the minimum required to complete our tests in the time
limit imposed by the scheduling system.

Practical implementations
for both polynomial and rational approximations require the calculation
of the lowest lying eigenmodes of $H_w$. We also need to compute the
eigenmodes of the overlap operator $D$ to speed up propagator calculation.
To compute these eigenmodes, we use implicitly restarted
Arnoldi factorization~\cite{Sorensen:1992fk,Lehoucq:1996}.
For a matrix $A\in \mathds{C}^{n\times n}$, if we desire $\ell$ eigenmodes, 
we construct an Arnoldi factorization~\cite{Arnoldi:1951}:
\beq
A V_k = V_k H_k + f_k e_k^\dagger \quad\text{with}\quad (e_k)_n=\delta_{k,n}\,,
\eeq
where $V_k=\{v_1,\dots,v_k\}\in\mathds{C}^{n\times k}$ satisfies 
$V_k^\dagger V_k=\iden_k$, i.e. the $n$-dimensional vectors on the columns
of $V_k$ are orthonormal. The matrix $H_k\in\mathds{C}^{k\times k}$ is an upper 
Hessenberg matrix that is the restriction of $A$ onto the Krylov space 
${\cal K}_k(A, v_1)$. The eigenvalues of $H_k$ represent Ritz estimates of the
eigenvalues of $A$ and the residue, $f_k$, can be used to gauge their accuracy.
In practical calculations $k\ll n$.
The implicitly restarted method uses a subspace of dimension $k$ significantly
larger than $\ell$, the desired number of eigenvectors. We set $k=\ell+\Delta\ell$, 
and at every step we remove from the
$k$ eigenmodes of $H_k$ the $\Delta\ell$ undesired ones. The 
factorization is restarted using the new starting subspace and the whole process
repeats until convergence. The optimal choice for our codes is 
$\Delta\ell \approx 1.5\ell$ and then the number of vectors
that need to be stored in memory is $2.5\ell$. The first iteration in this 
algorithm requires $k$ matrix multiplications and $k(k-1)/2$ orthogonalizations.
The subsequent iterations require only $\Delta\ell$ matrix-multiplications
and $k(k-1)/2-\ell(\ell-1)/2$ orthogonalizations.

\begin{figure}[t]
\centering
\includegraphics[width=3.2in]{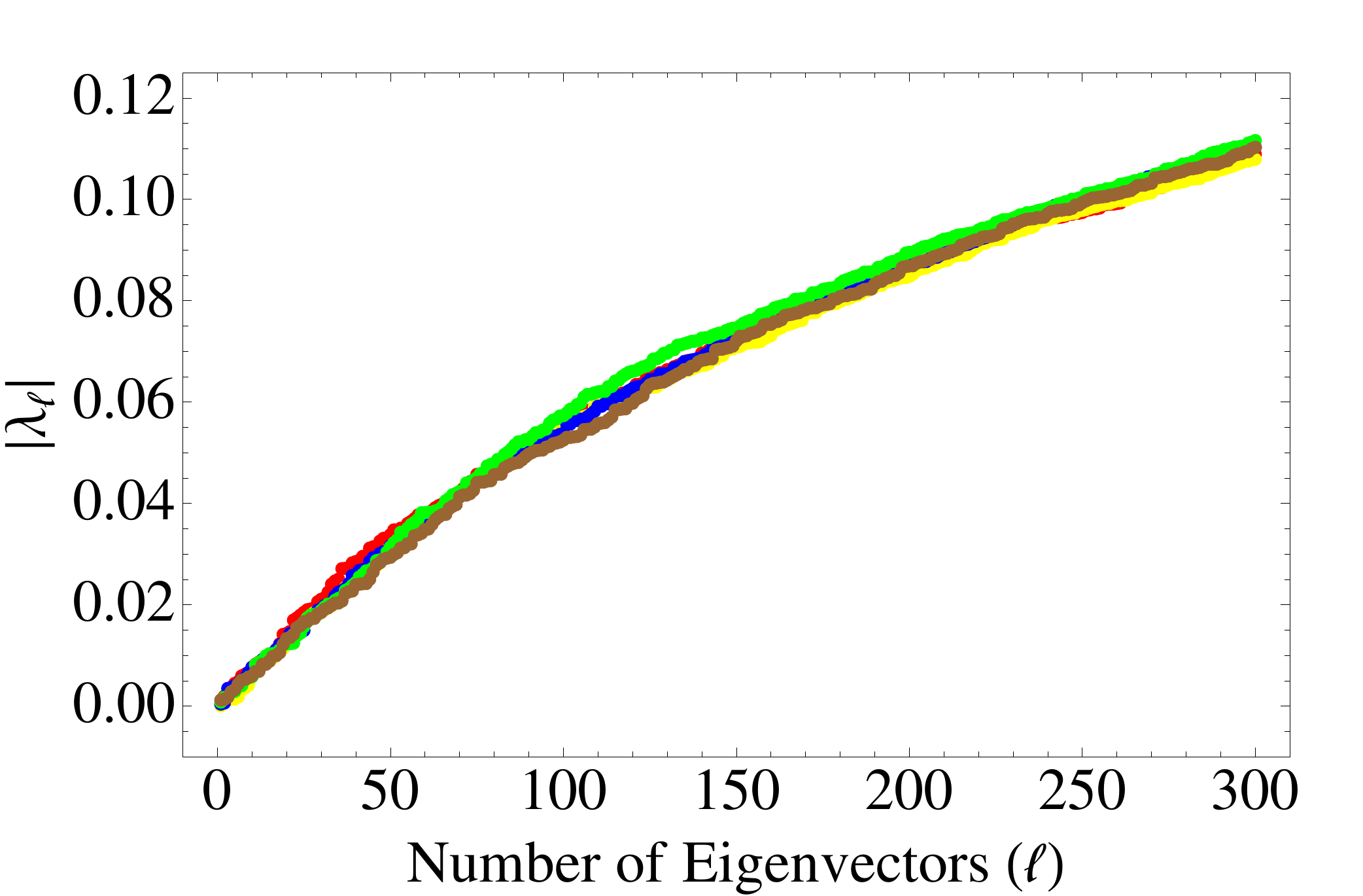}
\caption{Eigenvalue magnitude for $H_w$ as a function of eigenspace size. 
Each color represents a different $24^3\times 64$ lattice.}
\label{fig:hwilson-spect}
\end{figure}

\begin{figure}[t]
\centering
\includegraphics[width=3.2in]{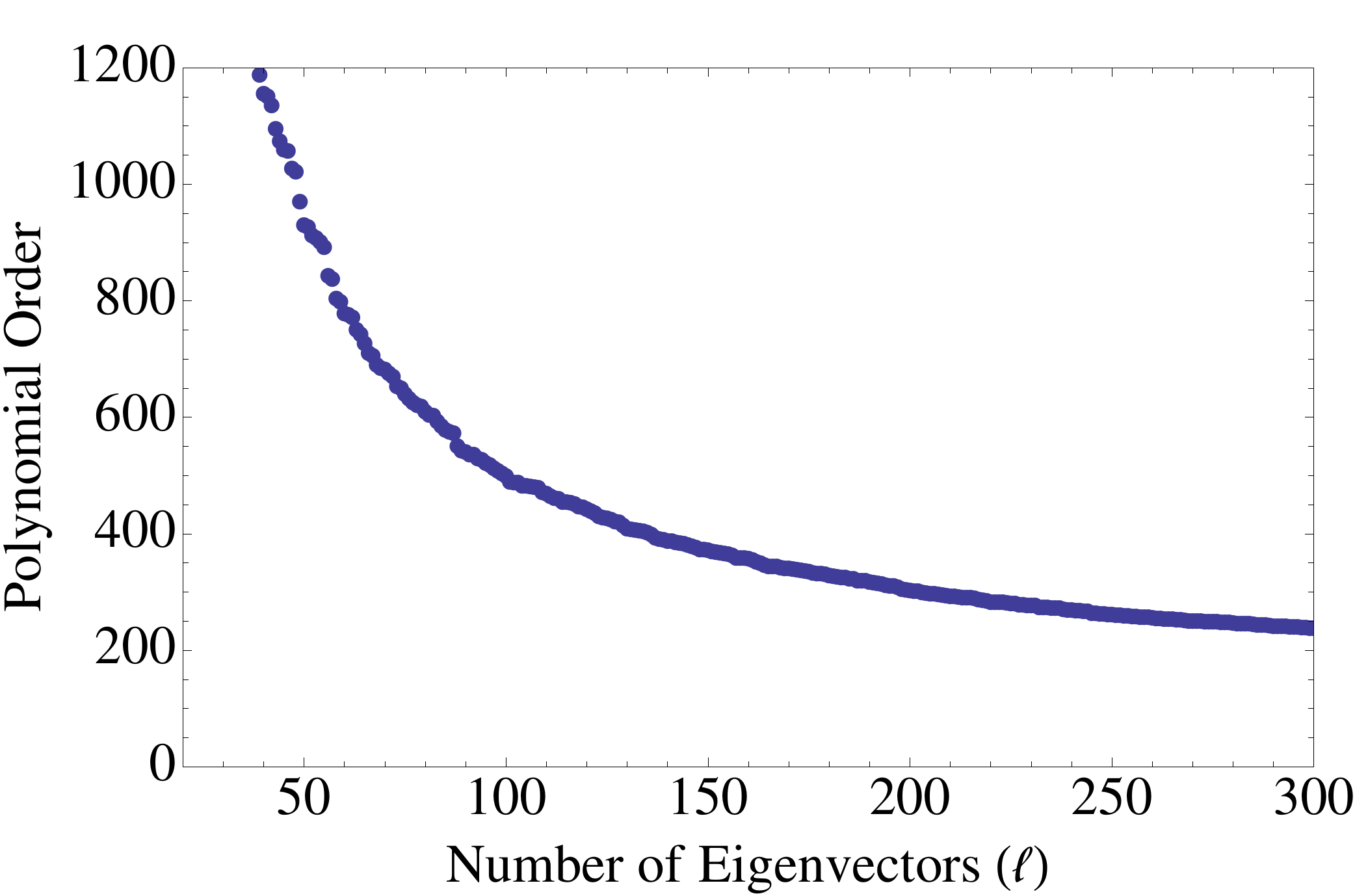}
\caption{Polynomial order required to approximate the sign function, $\sign(H_w)$ with a 
precision $\delta=10^{-9}$ in the subspace where $|\lambda|>|\lambda_\ell|$.}
\label{fig:poly-order}
\end{figure}

We now focus our discussion on the $H_w$ eigensolver. The number of desired
modes is dictated by both the structure of the low-lying spectrum of $H_w$ and
the available memory. In Fig.~\ref{fig:hwilson-spect} we plot the magnitude 
$|\lambda_\ell|$ as a function of $\ell$ for a handful of lattices from our
ensemble. It is clear that the spectral structure varies 
very little as we change the lattice. This can also be easily correlated with
the performance of the overlap operator routine: the most expensive part is the
sign multiplication routine, $\phi\gets Q P(Q^2)\psi$, which for polynomial
approximation is directly proportional to the order of the polynomial.
Using the empirical formula~\cite{Giusti:2002sm}
\beq
\delta = A e^{-b n\sqrt\epsilon},
\eeq
with $A=0.41$ and $b=2.1$ and setting 
$\sqrt\epsilon=|\lambda_\ell|/\lambda_\text{max}$ we can compute the order of
the polynomial required to achieve a precision of $\delta=10^{-9}$. The results
are shown in Fig.~\ref{fig:poly-order}. We see that going from $\ell=100$ to
$\ell=200$ the polynomial order is reduced by about a factor of two. Since
our GPU cluster can only store $500$ vectors in device memory we set $\ell=200$ 
(recall that the Arnoldi eigensolver uses $2.5\times \ell$ vectors).

To accelerate the convergence of $H_w$ eigenvectors we use Chebyshev
acceleration~\cite{Neff:2001zr}: we compute the eigenvectors of a
polynomial $T_n(H_w^2)$ that has a more suitable eigenvalue structure
but the same eigenvectors. We use a Chebyshev polynomial of the order
$100$ which speeds the convergence considerably (usually the Arnoldi
eigensolver converges in one iteration). Our GPU cluster needs
$0.27$ hours to converge whereas the Cray machine needs $0.60$ hours.
Thus, one GPU is equivalent to {\em 18 CPU cores} for this code.

Consumer level GPUs are significantly cheaper than the Tesla GPUs 
and offer similar performance for our codes. However, they have less
memory available, usually $1.5\text{--}2\GB$ per GPU. We are forced then to
use CPU memory to store the eigenvectors and the GPUs only to carry out
the {\tt dslash} multiplication. Due to the overhead associated with
moving the vector over the PCI bus the effective performance of {\tt dslash}
is reduced by a factor of 5 to 10. In our case this problem is less
severe because we use Chebyshev acceleration: computing $T_{100}(H_w^2)$
requires 200 {\tt dslash} multiplications and the overhead is paid only
once. Thus our effective {\tt dslash} performance is very close to the
pure GPU case. However, the orthogonalizations are computed on the CPU
in this case and this adds a significant overhead. This {\em mixed} code
takes $0.43$ hours to converge on our GPU cluster, 60\% more than the
pure GPU code. This ratio depends on the number of eigenvectors requested, $\ell$,
and it will become worse as $\ell$ is increased.
This is because the GPU time will increase linearly with $\ell$ since the GPUs are
responsible for the {\tt dslash} multiplications whereas
the CPU time increases quadratically since the number of orthogonalizations 
required increases quadratically with $\ell$.

We now turn our attention to the overlap operator $D$. As discussed in 
Section~\ref{sec:overlap}, the sign function can be approximated using
either a polynomial or a rational approximation. In the polynomial case,
we expand the polynomials in terms of Chebyshev polynomials and use
a Clenshaw recursion to evaluate $P(Q^2)\psi$. For the rational approximation
we expand the rational function:
\beq
P(Q^2)\psi = \sum_{i=1}^n \frac{b_i}{Q^2+c_i}\psi\,,
\eeq
and compute $(Q^2+c_i)^{-1}\psi$ for all $i$'s at once using a multi-shifted
CG method. The advantage of the polynomial approximation is
that the memory requirements are small. Clenshaw recursion needs only 
5 vectors whereas the multi-shifted CG needs $2n+3$ vectors. While the
rational approximation converges very fast we still need $n=12\text{--}20$.
The {\em double pass}~\cite{Neuberger:1998jk} variant of the 
rational approximation alleviates this problem at the cost of doubling
the number of {\tt dslash} matrix multiplications. In spite of this,
the double-pass algorithm is faster than the {\em single pass}
version~\cite{Chiu:2003ub} due to the reduced number of vector operations
required. Moreover, it was found that the double-pass algorithm takes
the same time irrespective of the order of the rational approximation.

\begin{figure}[t]
\centering
\includegraphics[width=3.3in]{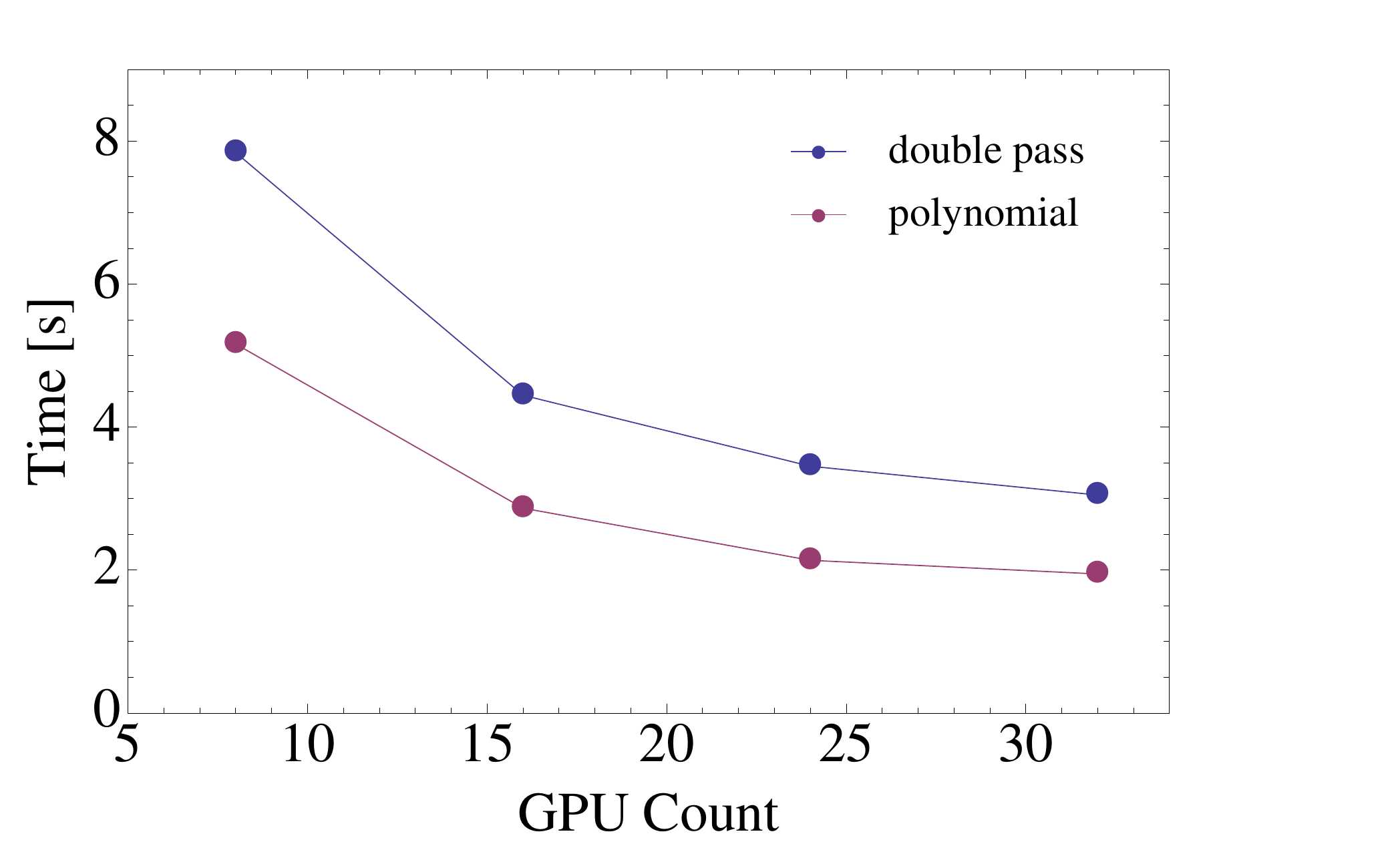}
\caption{Polynomial order required to approximate the sign function, $\sign(H_w)$ with a 
precision $\delta=10^{-9}$ in the subspace where $|\lambda|>|\lambda_\ell|$.}
\label{fig:overlap-comp}
\end{figure}

To decide on the optimal strategy, we compared the polynomial approximation
with the double-pass algorithm and we run the codes on $8$, $16$, $24$ and
$32$ GPUs. For this comparison we use $\ell=40$ to fit in the memory available
in $8$ GPU case. The double-pass algorithm used $n=18$ and the exit
criterion for CG was set to $\delta=10^{-10}$. The polynomial approximation
was tuned to the same precision.
The results of the test are presented in Fig.~\ref{fig:overlap-comp}. It is
clear that the polynomial approximation is the better choice and our codes
are based on it. When we use all 200 eigenvectors the overlap multiplication routine
requires $1.1$ seconds on 32 GPUs whereas our Cray machine need $3.3$ seconds. Thus,
one GPU is equivalent to {\em 24 CPU cores} for this routine.

We now turn our attention to the problem of computing the quark propagators.
To use deflation, we have to compute the overlap eigensystem using the Arnoldi method.
We first compute the eigenvectors of $\gamma_5 D$ in one chiral sector. 
Our $\gamma$-matrix basis is chiral and we can store the vectors
that have definite chirality using only half the storage required
for a regular vector. We can then store all $\ell=200$ $H_w$ eigenvectors
required for computing the overlap operator in device memory together
with the $2.5\times\ell'=250$ half-vectors used by the Arnoldi algorithm.
On our GPU cluster computing $\ell'=100$ eigenvectors to a precision of
$\delta=10^{-10}$ takes $2.7$ hours. On the Cray machine this takes
$10.6$ hours. Thus, one GPU is equivalent to {\em 26 CPU cores} for this code.

On systems with reduced memory we can move the Arnoldi half-vectors on the
CPU since they are accessed less frequently. In this case our GPU cluster
requires $4.0$ hours to converge which is 50\% more than in the pure GPU
case. 

We use an adaptive CG method~\cite{Cundy:2004pza}
to compute $D(m)^{-1}\psi$ with a precision of $10^{-8}$.
For a quark mass corresponding to $m_\pi\approx 200\MeV$ the adaptive
method is 60\% faster than the regular CG. 
To compute a full propagator for this mass the GPU cluster needs
$0.52$ hours and the Cray machine needs $2.3$ hours. One GPU is then
equivalent to {\em 35 CPU cores} for this code.

Overall, a quark propagator calculation takes $3.5$ hours on our
32 GPU cluster compared to $13.5$ hours on the 256 cores Cray machine.
This is consistent with the ratio of 22 CPU cores per one GPU that was
computed for the {\tt dslash} routine.

\section{Conclusions}

In this paper, we showed how to effectively employ GPUs for lattice
QCD calculations using the overlap operator. The most challenging aspect
of this calculation is the large amount of memory that needs to be accessed
frequently. To deal with this issue, we had to implement our codes
to run in parallel on multiple GPUs. 

Our optimization efforts focused on implementing the {\tt dslash} 
routine efficiently. For $24^3\times 64$ lattices our 
implementation scales reasonably well up to 32 GPUs where we still
run at 50\% efficiency. CPU clusters of comparable performance have
worse scaling efficiency and the GPU/CPU core ratio for similar performance
is even larger than 22, the ratio measured in the single-GPU case.

To compute the overlap quark propagators we need to implement eigensolvers
for both $H_w$ and $D$. We used the implicitly restarted Arnoldi algorithm, and
we found that the performance is very similar to the {\tt dslash} routine when
all vectors reside in device memory. On systems where the device memory is
not sufficient to hold all vectors, we found that storing the Arnoldi vectors in
CPU memory is a reasonable alternative. The performance penalty is only 50--60\%. 

We compared two different approximation strategies for the sign function used to
define the overlap operator. We find that the polynomial approximation is better
than the double-pass algorithm. Using this approximation the overlap operator
runs at a rate equivalent to 24 CPU cores. In the future, we plan to investigate the single pass algorithm.

The quark propagator is computed using an adaptive precision CG method which
runs at a rate equivalent to 35 CPU cores. Overall, the GPU/CPU performance
ratio for our codes is compatible with the ratio measured for the {\tt dslash}
routine. This result is not surprising
since the most time consuming part of these codes is the {\tt dslash}
routine, but it takes careful planning to work around all possible bottlenecks.


\section*{Acknowledgments}

This work is partially supported by DOE grant DE-FG02-95ER-40907. 
We wish to thank Mike Clark, Ron Babich and Balint Joo for useful discussions. 
The computational resources for this project were provided 
in part by the George Washington University IMPACT initiative.



\bibliographystyle{IEEEtran}
\bibliography{my-references}
%
%
%

\end{document}